\documentclass[11pt,twoside]{article}
\usepackage{asp2010}

\resetcounters

\begin{document}

\title{Discovery of Super-Thin Disks in Nearby Edge-on Spiral Galaxies}
\author{Andrew~Schechtman-Rook and Matthew~A.~Bershady
\affil{Department of Astronomy, University of Wisconsin,\\475 N. Charter
  Street, Madison, WI 53706, USA}}

\begin{abstract}
We report the identification of a super-thin disk (h$_{z}\sim60$ pc) in the
edge-on spiral galaxy NGC 891. This component is only apparent after we
perform a physically motivated attenuation correction, based on detailed
radiation transfer models, to our sub-arcsecond resolution near-infrared
imaging. In addition to the super-thin disk, we are also find several
structural features near the center of NGC 891, including an inner disk
truncation at $\sim$3 kpc. Inner disk truncations may be commonplace among massive
spiral galaxies, possibly due to the effects of instabilities, such as bars. Having successfully
demonstrated our methods, we are poised to apply them to a small sample of
nearby edge-on galaxies, consisting both of massive and low-mass spirals.
\end{abstract}

\section{Introduction}
Understanding the vertical disk structure of spiral galaxies is crucial for
learning about how galaxies form and evolve over time
\citep{Samland03,Moster10}, as well as producing an accurate estimate for the
baryon content of disks \citep{Bershady10a,Bershady10b}. In the Milky Way,
with survey measurements of many thousands of individual stars, disk structure
can be examined in great detail \citep[e.g.][]{Bovy12}. When viewed as if it
were an external galaxy, the Milky Way's light distribution can be fit with just three main disk components:
the thick disk, containing old stars (scaleheight h$_{z}\sim1$ kpc); the
thin disk, the most domininant light-weighted component (h$_{z}\sim300$ pc);
and a super-thin disk, consisting mostly of young stars \citep[h$_z\sim100$ pc,][]{vanDokkum94}. The latter component goes by many names in the
literature; it is frequently referred to as the `young' or `star-forming'
disk. We prefer the term `super-thin' as it carries no assumption about the age of the disk. 

The existence of both thin and thick disks in other spiral galaxies, while an
uncertainty for many years \citep{vanDokkum94}, has now been confirmed
\citep{Dalcanton02,Comeron11}. However, the ubiquitous presence of dust near
the midplane of these galaxies \citep[especially for massive ones most similar
to the Milky Way,][]{Dalcanton04} prevents easy study of any super-thin
components.

Here we present initial results from our ongoing work to map edge-on disk
structure, using a combination of
high-resolution near-infrared (NIR) imaging and advanced Monte Carlo radiation
transfer models to remove the effects of dust and probe vertical surface
brightness profiles all the way down to the midplane. Our method is
illustrated by its application to NGC 891, but is broadly applicable to all
edge-on galaxies.

\articlefigure[scale=0.45]{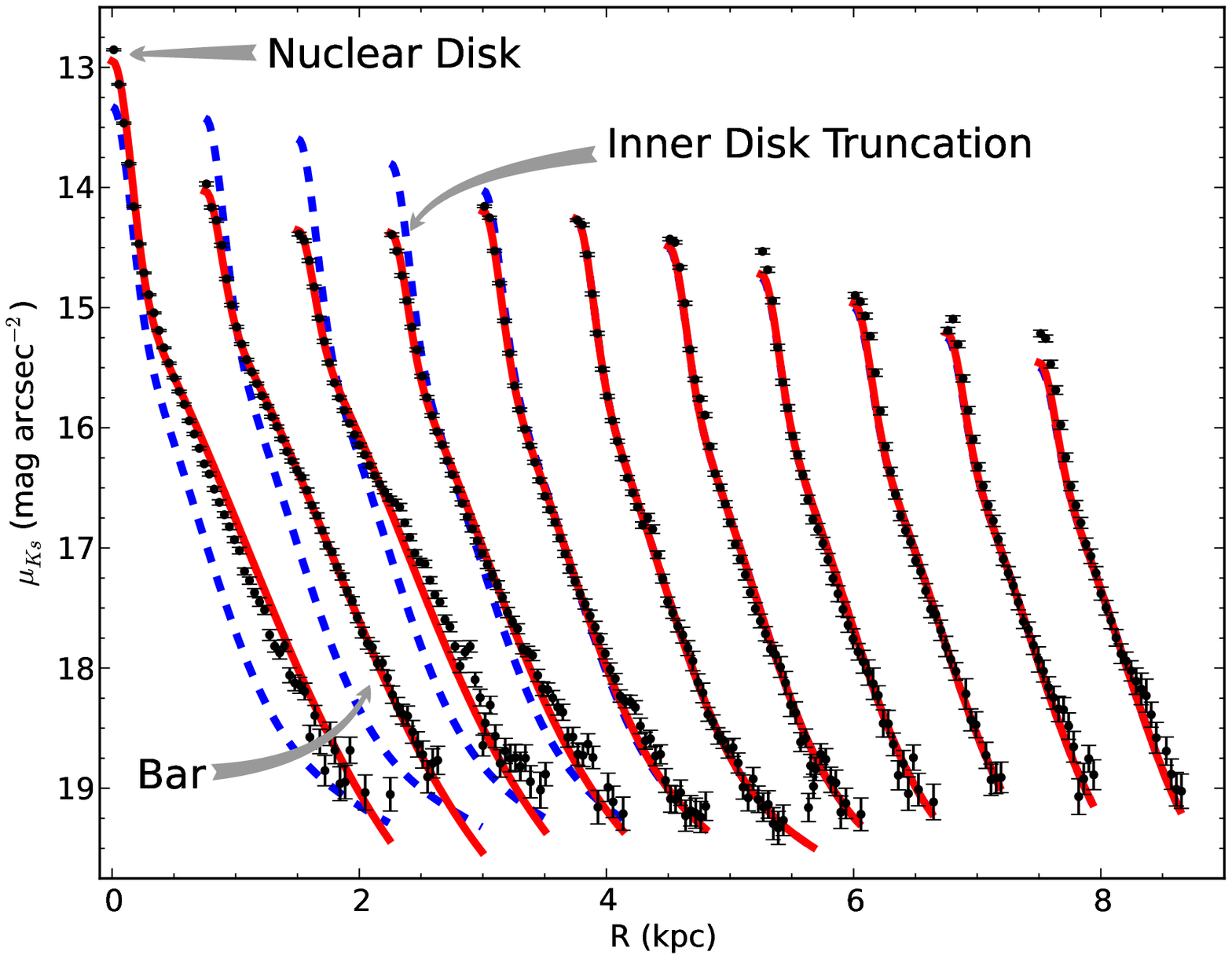}{fig:truncation}{Vertical
  profiles of NGC 891 at different radii. Data is shown as black dots, while
  dashed dark gray lines indicate the best fitting three-component disk model with
  exponential radial and vertical profiles. Inner regions deviating from this
  idealized model are marked by arrows indicating the additional structure
  required to fit the observed surface-brightness profiles. The light gray solid
  lines show our best fitting model including the inner disk truncation, bar,
  and nuclear disk. }
\section{Data and Model}
The following is a brief overview of our data and our radition transfer
modeling. Full details of our methodology, both for data reduction and producing our
attenuation correction, can be found in \citet{Schechtman-Rook13}.
\subsection{NIR Data}
Data were taken on the WIYN High-resolution InfraRed Camera
\citep[WHIRC;][]{Meixner10} on the WIYN 3.5 m telescope in 2011 October. NGC
891 is significantly larger in radius than the WHIRC field of view, so three
pointings were required to image out to just over $\pm$10 kpc in radius. Sky
images were interspersed with data exposures in order to compute an accurate
background estimate for such an extended object. The
images were processed using in-house reduction software, which performs all
aspects of data reduction from initial trimming of the images to creating the
final mosaic. 
\subsection{Attenuation Correction}
To compute the attenuation correction we first model NGC 891 using the
radiation transfer software HYPERION \citep{Robitaille11}; a model was created
which matched the integrated spectral energy distribution of NGC 891, and
images of this model with and without dust were compared to produce a
pixel-by-pixel attenuation map. The attenuation in a given pixel was
well-predicted by the model's K$_{\mathrm{s}}$-4.5$\mu$m color.
\articlefigure[scale=0.53]{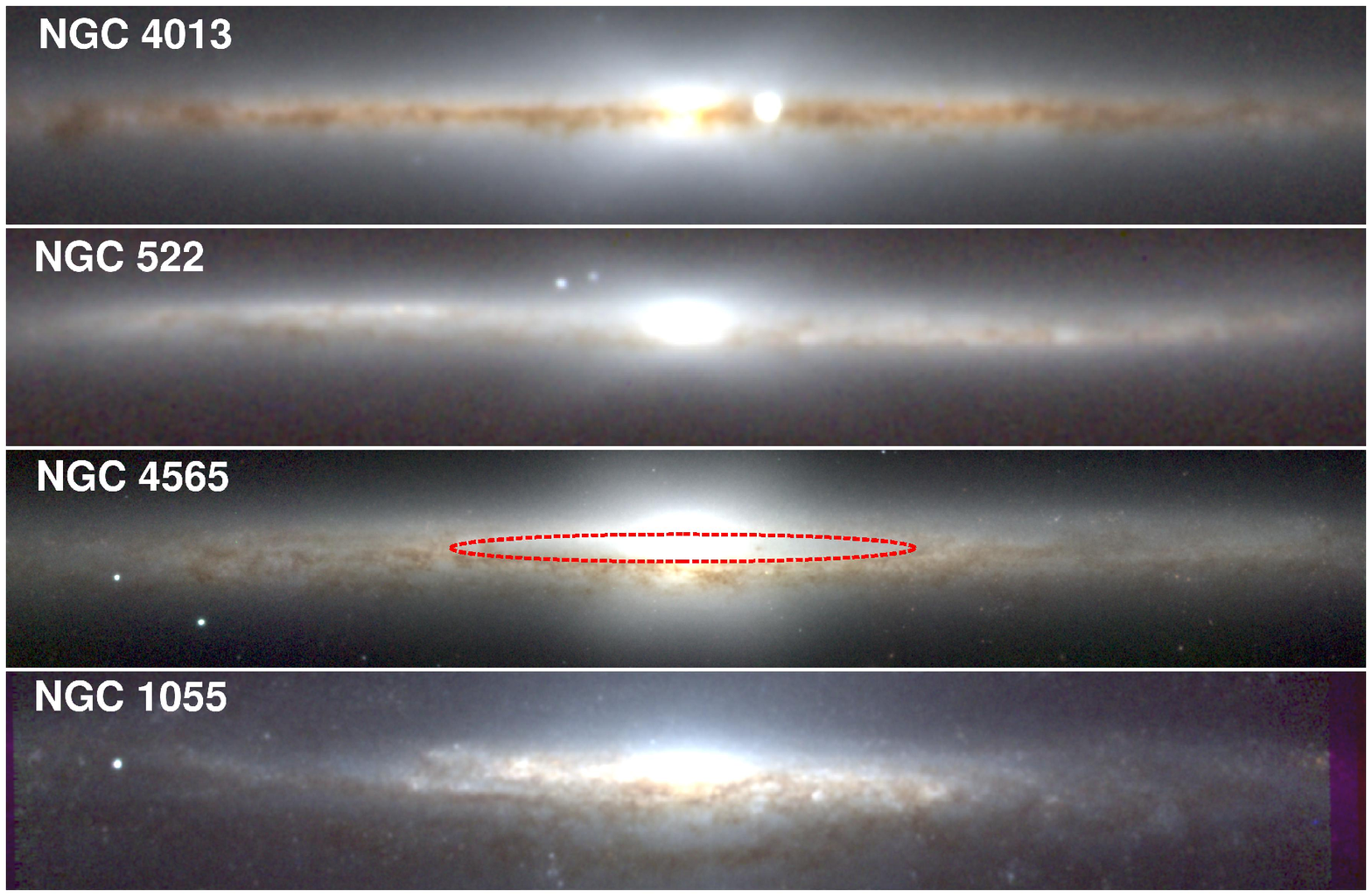}{fig:fastrotators}{JHK$_{\mathrm{s}}$
  false-color images of the isolated fast
  rotators (other than NGC 891) in our sample. The dashed ellipse
  indicates the inner truncation in visible in our images of NGC 4565.}

In order to use this color to correct our WHIRC data, we obtained an archival
{\it Spitzer} IRAC image of NGC 891. While this image is at significantly
lower resolution compared to our NIR data, the intrinsic smoothness of the luminosity
profile at 4.5$\mu$m allows us to preserve the resolution of our WHIRC data.

\section{Results}
\subsection{NGC 891}
Observed and attenuation-corrected JHK$_{\mathrm{s}}$ false-color images are
given in\\ \citet{Schechtman-Rook13}. We fit models with up to three discrete disk components, producing
best-fitting models using a Levenberg-Marquardt non-linear least-squares
fitter. For these initial models we ignored the central 3 kpc of NGC 891 to
avoid bulge/bar contamination. We find that three disks are necessary in order
to reproduce NGC 891's intrinsic K$_{\mathrm{s}}$-band surface-brightness at all
heights. These three components have very scale-heights to the Milky Way, with
the super-thin disk scale-height $h_{z}\approx$60 pc. 

We then investigated the ability of our models to fit the data within 3 kpc,
and found that models with just disks and R$^{1/4}$-law bulges were
  insufficient. In all, we had to add three new features to our best-fitting 3
  disk model: an inner disk truncation for all three disk components at R$\approx$3.1 kpc; a bar, which was
  well-approximated by an exponential disk with $h_{z}\approx$400 pc; and
  a nuclear extension of the super-thin disk, which has a scale-length of only
  250 pc. An R$^{1/4}$-law bulge is not necessary to reproduce the light
  profile of NGC 891, and actually results in significantly worse fits if used
  in place of the nuclear super-thin disk. The light from the bar is roughly
  equivalent to the light lost due to the inner disk truncation, implying that
  the bar is composed of stars swept up from NGC 891's disks. 

\articlefigure[scale=0.53]{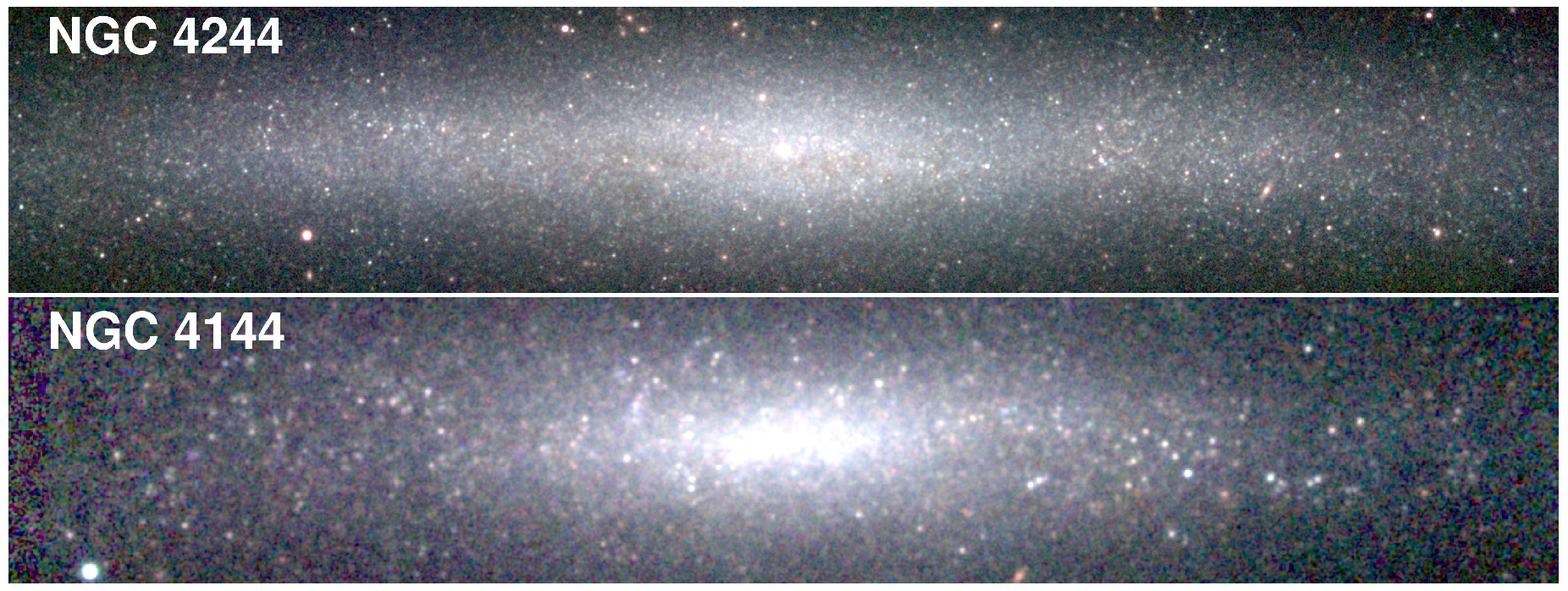}{fig:slowrotators}{JHK$_{\mathrm{s}}$
  false-color images of the two slow-rotators (V$<$90 km sec$^{-1}$) in our
  sample. These galaxies show little evidence for dust-lane attenuation but
  have extensive star-clusters throughout the vertical range of the disk.}
\subsection{Other Galaxies}
In addition to NGC 891, we have a small sample of nearby edge-on galaxies
suitable for this analysis. We generally selected systems with properties
similar to the Milky Way and NGC 891 (Figure \ref{fig:fastrotators}); however
we have also included some low-mass spirals (Figure \ref{fig:slowrotators}) as well as a galaxy undergoing a
minor interaction (bottom panel of Figure \ref{fig:fastrotators}). In both
figures the images are not attenuation corrected.

Of special note is NGC~4565 (second from bottom panel of Figure \ref{fig:fastrotators}),
which has a clearly visible inner disk truncation. Interestingly this also
appears to be the situation in the Milky Way \citep[at least in dust
emission, see][]{Robitaille12}. While three galaxies is too small a sample to draw
definitive conclusions, it appears inner disk truncations may be a common
feature of massive spiral galaxies. 

\acknowledgements This research was supported by NSF AST-1009471.

\bibliography{schechtman-rooka}

\begin{thebibliography}{}
\expandafter\ifx\csname natexlab\endcsname\relax\def\natexlab#1{#1}\fi
\expandafter\ifx\csname url\endcsname\relax
  \def\url#1{\texttt{#1}}\fi
\expandafter\ifx\csname urlprefix\endcsname\relax\def\urlprefix{URL }\fi
\providecommand{\eprint}[2][]{\url{#2}}

\bibitem[{{Bershady} et~al.(2010{\natexlab{a}}){Bershady}, {Verheijen},
  {Swaters}, {Andersen}, {Westfall}, \& {Martinsson}}]{Bershady10a}
{Bershady}, M.~A., {Verheijen}, M.~A.~W., {Swaters}, R.~A., {Andersen}, D.~R.,
  {Westfall}, K.~B., \& {Martinsson}, T. 2010{\natexlab{a}}, \apj, 716, 198.
  \eprint{1004.4816}

\bibitem[{{Bershady} et~al.(2010{\natexlab{b}}){Bershady}, {Verheijen},
  {Westfall}, {Andersen}, {Swaters}, \& {Martinsson}}]{Bershady10b}
{Bershady}, M.~A., {Verheijen}, M.~A.~W., {Westfall}, K.~B., {Andersen}, D.~R.,
  {Swaters}, R.~A., \& {Martinsson}, T. 2010{\natexlab{b}}, \apj, 716, 234.
  \eprint{1004.5043}

\bibitem[{{Bovy} et~al.(2012){Bovy}, {Rix}, {Liu}, {Hogg}, {Beers}, \&
  {Lee}}]{Bovy12}
{Bovy}, J., {Rix}, H.-W., {Liu}, C., {Hogg}, D.~W., {Beers}, T.~C., \& {Lee},
  Y.~S. 2012, \apj, 753, 148. \eprint{1111.1724}

\bibitem[{{Comer{\'o}n} et~al.(2011){Comer{\'o}n}, {Elmegreen}, {Knapen},
  {Salo}, {Laurikainen}, {Laine}, {Athanassoula}, {Bosma}, {Sheth}, {Regan},
  {Hinz}, {Gil de Paz}, {Men{\'e}ndez-Delmestre}, {Mizusawa},
  {Mu{\~n}oz-Mateos}, {Seibert}, {Kim}, {Elmegreen}, {Gadotti}, {Ho},
  {Holwerda}, {Lappalainen}, {Schinnerer}, \& {Skibba}}]{Comeron11}
{Comer{\'o}n}, S., {Elmegreen}, B.~G., {Knapen}, J.~H., {Salo}, H.,
  {Laurikainen}, E., {Laine}, J., {Athanassoula}, E., {Bosma}, A., {Sheth}, K.,
  {Regan}, M.~W., {Hinz}, J.~L., {Gil de Paz}, A., {Men{\'e}ndez-Delmestre},
  K., {Mizusawa}, T., {Mu{\~n}oz-Mateos}, J.-C., {Seibert}, M., {Kim}, T.,
  {Elmegreen}, D.~M., {Gadotti}, D.~A., {Ho}, L.~C., {Holwerda}, B.~W.,
  {Lappalainen}, J., {Schinnerer}, E., \& {Skibba}, R. 2011, \apj, 741, 28.
  \eprint{1108.0037}

\bibitem[{{Dalcanton} \& {Bernstein}(2002)}]{Dalcanton02}
{Dalcanton}, J.~J., \& {Bernstein}, R.~A. 2002, \aj, 124, 1328.
  \eprint{arXiv:astro-ph/0207221}

\bibitem[{{Dalcanton} et~al.(2004){Dalcanton}, {Yoachim}, \&
  {Bernstein}}]{Dalcanton04}
{Dalcanton}, J.~J., {Yoachim}, P., \& {Bernstein}, R.~A. 2004, \apj, 608, 189.
  \eprint{arXiv:astro-ph/0402472}

\bibitem[{{Meixner} et~al.(2010){Meixner}, {Smee}, {Doering}, {Barkhouser},
  {Miller}, {Orndorff}, {Knezek}, {Churchwell}, {Scharfstein}, {Percival},
  {Mills}, {Corson}, \& {Joyce}}]{Meixner10}
{Meixner}, M., {Smee}, S., {Doering}, R.~L., {Barkhouser}, R.~H., {Miller}, T.,
  {Orndorff}, J., {Knezek}, P., {Churchwell}, E., {Scharfstein}, G.,
  {Percival}, J.~W., {Mills}, D., {Corson}, C., \& {Joyce}, R.~R. 2010, \pasp,
  122, 451

\bibitem[{{Moster} et~al.(2010){Moster}, {Macci{\`o}}, {Somerville},
  {Johansson}, \& {Naab}}]{Moster10}
{Moster}, B.~P., {Macci{\`o}}, A.~V., {Somerville}, R.~S., {Johansson}, P.~H.,
  \& {Naab}, T. 2010, \mnras, 403, 1009. \eprint{0906.0764}

\bibitem[{{Robitaille}(2011)}]{Robitaille11}
{Robitaille}, T.~P. 2011, \aap, 536, A79. \eprint{1112.1071}

\bibitem[{{Robitaille} et~al.(2012){Robitaille}, {Churchwell}, {Benjamin},
  {Whitney}, {Wood}, {Babler}, \& {Meade}}]{Robitaille12}
{Robitaille}, T.~P., {Churchwell}, E., {Benjamin}, R.~A., {Whitney}, B.~A.,
  {Wood}, K., {Babler}, B.~L., \& {Meade}, M.~R. 2012, \aap, 545, A39.
  \eprint{1208.4606}

\bibitem[{{Samland} \& {Gerhard}(2003)}]{Samland03}
{Samland}, M., \& {Gerhard}, O.~E. 2003, \aap, 399, 961.
  \eprint{arXiv:astro-ph/0301499}

\bibitem[{{Schechtman-Rook} \& {Bershady}(2013)}]{Schechtman-Rook13}
{Schechtman-Rook}, A., \& {Bershady}, M.~A. 2013, \apj, 773, 45.
  \eprint{1306.6076}

\bibitem[{{van Dokkum} et~al.(1994){van Dokkum}, {Peletier}, {de Grijs}, \&
  {Balcells}}]{vanDokkum94}
{van Dokkum}, P.~G., {Peletier}, R.~F., {de Grijs}, R., \& {Balcells}, M. 1994,
  \aap, 286, 415

\end{thebibliography}

\end{document}